\documentclass[12pt]{article}
\usepackage{amsthm,amssymb,amsmath}

\renewcommand{\d}{\operatorname{d}}\usepackage[british]{babel}

\begin{document}

\title{On  Twistor Solutions of the dKP Equation\thanks{Partially supported by the DGESIC. Proyecto
PB98-0821.} }

\author{
Francisco Guil$^1$, Manuel Ma{\~n}as$^2$ and Luis Mart{\'\i}nez
Alonso$^3$ \\
 \textsl{Departamento de F{\'\i}sica Te{\'o}rica II},
\textsl{Facultad de Ciencias F{\'\i}sicas}, \\
\textsl{Universidad Complutense, 28040 Madrid, Spain}\\
\small{\texttt{$^1$fguil@fis.ucm.es,
$^2$manuel@darboux.fis.ucm.es, $^3$luism@fis.ucm.es}}
   }

\maketitle

\newtheorem{definition}{Definition}[section]

\newtheorem{theorem}{Theorem}[section]

\newtheorem{example}{Example}[section]

\numberwithin{equation}{section}

\begin{abstract}
The factorization problem for the group of canonical
transformations  close to the identity and the corresponding
 twistor equations for an ample family of canonical variables
are considered. A method to deal with these reductions is
developed for the construction classes of nontrivial solutions of
the dKP equation.
\end{abstract}



\section{Introduction}\label{sec:int}

This paper deals with the integration problem of the
dispersionless Kadomtsev--Petviashvili (dKP) equation, the first
member of the integrable hierarchy obtained from the ordinary KP
hierarchy through the dispersionless limit. That limit coincides
with the quasi-classical approximation for the underlying quantum
space of differential operators of the KP equation which, for the
dKP case,  becomes the classical phase space endowed with a
Poisson structure.

The theory and applications of the nonlinear models arising in the
dispersionless (or quasiclasssical) limits of the integrable
systems of KdV-type have been  active subjects of research for
more than twenty years (see for example \cite{1}-\cite{14}).
However, the theory of their solution methods seems far from being
completed. Indeed, only for a few cases \cite{15}-\cite{17} the
dispersionless limit of the inverse scattering method is
available and \emph{dispersionless versions} of ordinary direct
methods like the $\overline{\partial}$-method are not yet fully
developed \cite{18}.

 In \cite{3}-\cite{4} Kodama and Gibbons provided a direct method
for finding solutions of the dispersionless KP (dKP) equation and
its associated dKP hierarchy of nonlinear systems. The main
ingredient of their method is the use of reductions of the dKP
hierarchy formulated in terms of hydrodynamic-type equations. An
alternative direct method for solving the dKP hierarchy from its
reductions was recently proposed in \cite{19} . It is based on the
characterization of reductions ((and hodograph solutions)) of the
dKP hierarchy by means of certain systems of first-order partial
differential equations.
 In \cite{5} Takasaky and Takebe showed that the factorization problem for
the group of canonical transformations in two-dimensional phase
space provides  a direct solution method for the dKP equation.
Furthermore, that method has a twistor interpretation. The group
of canonical transformations acts on the phase space, viewed as a
part of the Lie algebra of that group, through the adjoint
representation. Hence, the factorization problem in the group is
represented in the phase space too. That representation is called
the twistor formulation of the factorization problem and admits
several geometric interpretations. This twistor formulation allows
for the use of the canonical formalism of Classical Mechanics in
the resolution of the factorization problem for solving the dKP
equation. The canonical formalism enters in the form  of
generating functions for canonical transformations. When expressed
in terms of appropriate variables, these generating functions
provide us with a trivialization of the group structure in the
sense that transform right-derivatives in the group of canonical
transformations into ordinary derivatives. This seems to be a
relevant fact in the whole integration scheme.

The Hamilton--Jacobi method of integration of the canonical
equations appears in the present context of multi-time Hamiltonian
formalism as a trivialization, again, of the group structure. The
zero-curvature condition, which holds for the right-differential
of a function in the group, proves to be equivalent to the
requirement that the Poincar{\'e}--Cartan action 1-form associated
to that right-differential be a closed differential form. The
dependence of the momentum in coordinates and time, the main
feature of the Hamilton--Jacobi theory, compensates the
non-commutative Poisson structure of the phase space and
transforms the zero-curvature condition in an equivalent closed
differential 1-form which is the differential of the mechanical
action of the system.

In this paper a method for solving the dKP equation based in the
resolution of the twistor equations is given. It is concerned with
a class of canonical transformations which can be considered as
defining the initial conditions for the factorization problem.
One of the main ingredients in our construction lays in the
observation that, regarding the twistor equations, negative
powers in the momentum variable can be substituted by negative
powers in the Lax function. This equivalence furnishes a method
that allows for the reduction of the twistor equations, in the
class of the chosen canonical variables, to a system of
first-order ordinary differential equations for a finite number
of the coefficients of a generating function; however, it reduces
to an algebraic system in the simplest cases.

Our main result consists in the characterization and construction
of a class of canonical transformations, the initial conditions
for the factorization problem or twistor data in the terminology
of \cite{5}, for which the twistor equations can be solved
explicitly. Through such resolution we obtain the corresponding
solutions of the dKP equation which are described by means of the
generating function of the canonical transformation under
consideration. The generating functions are characterized as
series of fractionary powers in the momentum variable.

The organization and content of the paper are as follows. Next, in
\S 2 we revisit the twistor equations from the factorization
problem point of view. We first present the factorization problem
in the group of canonical transformations close to the identity in
a two-dimensional phase space. Then, the twistor equations are
derived. We continue by emphasizing the role of generating
functions  in the factorization problem and the relevance of the
expansions in negative powers of the Lax function. At the end of
the section  some of the symmetries of the dKP are rederived. In
\S 3 we consider the general case characterized by canonical
variables $X$ and $P$ of finite order at $p=\infty$. In
particular, generating functions for these reductions are found
and explicit solutions of the dKP equation are constructed.

\section{The twistor equations}\label{sec:fac}


\subsection{The factorization problem}

Let \( \mathfrak{g} \) be the Lie algebra of functions \( F(p,x)
\) on the two complex variables \( p\) and  \( x\) given as the
sum of two subalgebras \( \mathfrak{g}=\mathfrak{g_{+}}\oplus
\mathfrak{g_{-}} \) which are the spaces of analytic functions on
the variable \( p \) at \( p=0 \) and \( p=\infty \),
respectively, with a common domain of definition. Functions in
\(\mathfrak{g_{-}} \) vanish at \( p=\infty \) and the commutator
in \( \mathfrak{g} \) is defined by the Poisson bracket
\[
 \{F_{1}, F_{2}\}=\frac{\partial F_{1}}{\partial p}  \frac{\partial F_{2}}
  {\partial x} -\frac{\partial F_{1}}{\partial x} \frac{\partial
  F_{2}}{\partial p}.
\]

Let \( L \) be an element in \( \mathfrak{g} \) that depends on
new variables \( t_{2}, t_{3}, \ldots  \) and has  the prescribed
form
\[
    L(p, x, t_{2}, t_{3}, \ldots)=p+\sum_{j\geq 1} u_{j}( x, t_{2},
    t_{3}, \ldots) p^{-j}.
\]
As usual, define the functions \( P_{n} \), \( n=2,3,\ldots \) as
the projections on \( \mathfrak{g_{+}} \) of the positive powers
of \( L \), \( P_{n} =L^{n}\mid_{+} \) in terms of which the
Lax--Sato hierarchy is given by the infinite set of equations
\[
   \frac{\partial L}{\partial t_{n}}=\{ P_{n}, L \},
\]
for \( n=2,3,\ldots \) that in particular imply the dKP
 equation for the function
\( u=2 u_{1} \)
\[
    \big(u_{t}-\frac{3}{2} u u_{x}\big)_{x} =\frac{3}{4} u_{yy}
\]
where \( u_{1} \) is the coefficient of \( p^{-1} \) in \( L \) and
we have set \( t_{2}=y, t_{3}=t \).

The Lax--Sato equations provide, in their first and more usual
interpretation, the compatibility conditions for the vanishing of
the curvature of the connection
\[
    \omega_{+}:=\sum_{n\geq 2} P_{n} \d t_{{n}}.
\]
For that differential 1-form one finds
\[
    \d\omega_{+}=\frac{1}{2} \{\omega_{+},\omega_{+} \}=\sum_{m < n}
    \{P_{m},P_{n} \} \d t_{m} \wedge \d t_{n}
\]
where the dKP equation is represented by
\(\partial_{3}P_{2}-\partial_{2}P_{3}+\{P_{2},P_{3} \} =0 \). We
can think also of the previous system as a collection of
Hamiltonian flows in the phase space with canonical coordinates \(
(p, x) \). There is an infinite set of commuting flows defined by
the Hamiltonians \( H_{n}=-P_{n} \), \( n=2,3,\ldots \)
\[
   \frac{\partial x}{\partial t_{n}} =\frac{\partial H_{n}}{\partial
   p},\quad
   \frac{\partial p}{\partial t_{n}} =-\frac{\partial H_{n}}{\partial
   x},
\]
whose compatibility conditions are precisely the equations of the dKP
hierarchy for the potentials in the function \( L \).

The structure underlying the Lax--Sato system can be conveniently
understood by means of the group of canonical transformations
connected with the identity. A transformation \( ( p, x)
\rightarrow (P, X) \) is canonical
 ($\d P\wedge \d X=\d p\wedge \d x$) and connected with the identity
if its of the form
\begin{align*}
     P & =  \exp(\mathrm{ad} K). p  =   p-K_{x}-\frac{1}{2}\{K,K_{x}\}-\cdots,  \\
     X & =  \exp(\mathrm{ad} K). x  =
     x+K_{p}+\frac{1}{2}\{K,K_{p}\}+\cdots,
\end{align*}
where $K=K(p,x)$ is a function in the phase space and
\[
K_x:=\frac{\partial K}{\partial x},\quad K_p:=\frac{\partial
K}{\partial p}.
\]

These formulas admit also an interpretation in terms of the
right-derivatives of a group element in the local Lie group \( G
\) defined by the Lie algebra \( \mathfrak{g} \). Let \( G=\exp
\mathfrak{g}  \) be the Lie group of canonical transformations
connected with the identity. Then,
 for the action of an element \( k=\exp K \)  on the canonical coordinates
\( (p, x) \) we obtain the expressions
\begin{align*}
        P  = & \mathrm{Ad}k. p  =  \exp(\mathrm{ad} K). p=
        p-k_{x}k^{-1},   \\
        X  = & \mathrm{Ad}k. x  =  \exp(\mathrm{ad} K). x  =
    x+k_{p}k^{-1}.
\end{align*}
Where we have used the right--differential
\[
\d k\;k^{-1}=\sum_{n\geq 0}\frac{1}{(n+1)!}(\mathrm{ad}K)^n\d K.
\]

The relevance of this group in the theory of the dKP equation
appears in connection with the formulation of the Lax--Sato system
as a factorization problem. Let \( G_{\pm}=\exp \mathfrak{g}_{\pm} \) be
the local Lie groups with Lie algebras \( \mathfrak{g}_{\pm}  \),
respectively, viewed as subgroups of the group \( G \). Let \(
t(p) \) be an element in \( \mathfrak{g}_{+}  \) of the form \[
t(p):=t_{2} p^{2}+t_{3} p^{3}+\cdots \] and define the canonical
transformation \( \psi= (\exp t(p)) k \). The factorization
problem in the group \( G \) with respect to the subgroups \( G_{\pm}
\) is then the equation \( \psi=  \psi_{-}^{-1} \psi_{+} \) for
the representation of a given element \(  \psi \) as the product
of a pair of group elements \(  \psi_{\pm} \) in the subgroups  \( G_{\pm}
\) respectively. In particular, for the element \( \psi \) defined
above we have the equation
\begin{equation}
      \mathrm{e}^{t(p)} k =  \psi_{-}^{-1} \psi_{+}
    \label{eq:fac1}
\end{equation}
the solutions of which depend both on the time variables \( \{t_{2},
t_{3},\ldots\}  \) and on the initial condition given by the constant
element \( k \). More precisely, the element \( k \) enters in the
solutions \(  \psi_{\pm} \) as the representative of a point in the
double coset space \( \Gamma_{-} \backslash G/ G_{+}\) where \(  \Gamma_{-}
\) is the centralizer of \( \exp  t(p) \) in \( G_{-} \).

In order to construct the Lax--Sato equations through the
factorization problem we take the right-differential with respect
to the time variables in  \eqref{eq:fac1} from which we get
\begin{equation}
    \d \psi_{-} \psi_{-}^{-1} +\mathrm{Ad} \psi_{-}.\d t(p) = \d \psi_{+}
    \psi_{+}^{-1}.
    \label{eq:fac1.1}
\end{equation}

 If we identify the function \( L \) in \( \mathfrak{g} \) with a
 point in
the orbit of \( p \) for the action of \( G_{-} \), \( L =
\mathrm{Ad} \psi_{-}. p \), the projection on \( \mathfrak{g}_{+}
\) of
\begin{equation}
     \mathrm{Ad} \psi_{-}.\d t(p)=\sum_{n\geq 2}  \mathrm{Ad} \psi_{-}
    p^{n} \d t_{n}= \sum_{n\geq 2}  L^{n} \d t_{n}
    \label{eq:fac1.2}
\end{equation}
implies that the differential form \( \omega_{+}=\d \psi_{+}
\psi_{+}^{-1}=\sum_{n\geq 2}  L^{n}|_{+} \d t_{n} \) is of zero
curvature and hence the dKP equation obtains from the
factorization problem. The meaning of these equations is that
positive and negative projections of the form \[ \omega :=
\sum_{n\geq 2}  L^{n} \d t_{n} \] in \eqref{eq:fac1.2}, \( \omega
= \omega_{+} - \omega_{-} \), are of zero-curvature, \(\omega_{\pm} =
\d \psi_{\pm} \psi_{\pm}^{-1} \), as follows from \eqref{eq:fac1.1}.

\subsection{Twistor equations}

The canonical formalism allows for a reformulation of the
factorization problem (\ref{eq:fac1}) in the twistor language
\cite{5}. Let \( (P,X) \) be new canonical variables defined by
the element \( k \) in (\ref{eq:fac1}), \( P = \mathrm{Ad}k. p, X
= \mathrm{Ad}k. x \), and define in addition the canonical pair \(
(L, M) \) through the action of the canonical transformation \(
\psi_{-} \exp t(p) \),
\begin{equation}
    L = \mathrm{Ad} ( \psi_{-} \mathrm{e}^{t(p)}). p, \quad
    M = \mathrm{Ad} ( \psi_{-} \mathrm{e}^{t(p)}). x.
    \label{eq:fac2}
\end{equation}
For these variables, that are constants of motion for the
Hamiltonian flows with Hamiltonians \( H_{n}=-P_{n}=L^{n}|_{+} \),
\( n=2,3,\ldots \):
\[
\frac{\partial L}{\partial t_{n}}+\{ H_{n}, L \}=0, \quad
\frac{\partial M}{\partial t_{n}}+\{ H_{n}, M \}=0,
\]
one finds the expressions
\begin{equation}
    L = p-\frac{\partial \psi_{-}}{\partial x} \psi_{-}^{-1},\quad
    M = \frac{\partial t(L)}{\partial L} + x +
    \frac{\partial \psi_{-}}{\partial p} \psi_{-}^{-1}.
    \label{eq:fac3}
\end{equation}
As a consequence of formula (\ref{eq:fac1}) we obtain the relation
\[
    \psi_{-} \mathrm{e}^{t(p)} k = \psi_{+}
\]
whose action on the pair of canonical variables \( (p, x) \) results in
an equivalent description of the factorization problem, namely
\[
    P(L, M) = \mathrm{Ad} \psi_{+}.p, \quad X(L, M) = \mathrm{Ad}
    \psi_{+}.x
\]
which finally lead to the equations for the function \( L \)
\begin{equation}
    P(L, M)|_{-} = 0, \quad X(L, M)|_{-} = 0
    \label{eq:fac4}
\end{equation}
that represent the twistor form of the factorization problem
\eqref{eq:fac1}.

\subsection{Generating functions}\label{sec:gen}

The most effective method to deal with canonical transformations
is furnished by the formulation in terms of their generating
functions. If we want to describe the canonical variables \( (L,
M) \) of \eqref{eq:fac3} it proves to be convenient to define a
generating function for this transformation \( \Phi(L,x) \) such
that the differential of it is given by \( \d\Phi(L,x) = M \d L+p
\d x \). For such a transformation one finds the expression
\begin{equation}
    \Phi(L,x) = x L + t(L) + \phi(L, x)
    \label{eq:gen1}
\end{equation}
where \( t(L):=t_2L^2+t_3L^3+\cdots \)  and \( \phi(L, x) \) is a
negative power series in \( L \), \( \phi(L, x)=\sum_{n\geq 1}
\phi_{n}(x) L^{-n}  \). With this definition we deduce the
relations
\begin{equation}
    p =\frac{\partial \Phi}{\partial x} = L + \frac{\partial \phi}{\partial x}
    \label{eq:gen2}
\end{equation}
and
\begin{equation}
    M = \frac{\partial \Phi}{\partial L} = x + \frac{\partial
    t}{\partial L} + \frac{\partial \phi}{\partial L}.
    \label{eq:gen3}
\end{equation}
These formulas are to be compared with the corresponding expressions
for \( L \) and \( M \) as given by \eqref{eq:fac3} that imply the
relations
\begin{equation}
    \frac{\partial \phi}{\partial x} (L, x) =
    \frac{\partial \psi_{-}}{\partial x} \psi_{-}^{-1} (p, x),
    \hspace{.3in}
    \frac{\partial \phi}{\partial L} (L, x) =
    \frac{\partial \psi_{-}}{\partial p} \psi_{-}^{-1} (p, x)
    \label{eq:gen3.1}
\end{equation}
provided \( p \) and \( L \) satisfy \( p = L + \phi_{x} \) as in
\eqref{eq:gen2}.

From these equations we learn that functions \( \phi \) and \(
\psi_{-} \) are equivalent negative power series to describe the
solutions of the factorization problem. The main advantage of \(
\phi \) as compared with \( \psi_{-} \) is that it allows for the
use of ordinary partial derivatives instead the  right-invariant
derivatives necessary when dealing with \( \psi_{-} \). Such a
simplification is achieved by choosing the mixed independent
variables \( (L,x) \) for the generating function \( \Phi \) of
the canonical transformation. Greater simplification also is
gained in describing the coefficients of \( \phi (L, x) \), for if
we represent \( \psi_{-} = \exp \Psi \) in terms of a negative
power series \( \Psi = \sum_{j\geq 1} \Psi_{j} p^{-j} \), the
right-differential \( \omega_{-} = \d \psi_{-} \psi_{-}^{-1} \) is
then
\[
    \omega_{-} = \d \Psi + \frac{1}{2} \{\Psi, \d \Psi\} + \cdots.
\]
Integrating this equation along a conveniently chosen closed path
gives
\[
   \d \Psi_{1} = \frac{1}{2 \pi i} \int \omega_{-}(p) \d p,
\]
from which we deduce, taking into account \eqref{eq:fac1.1} and the
definition of \( \omega \), the relation
\[
   \d \Psi_{1} = - \frac{1}{2 \pi i} \int \omega (p) \d p = \frac{1}{2 \pi
   i} \int \omega (L) \sum_{k\geq 1} k L^{-k-1} \phi_{k x} \d L
\]
after we change the integration variable according to formula
\eqref{eq:gen2}. Upon substitution of \( \omega (L) = \sum_{j\geq
2}L^{j} dt_{j} \) we finally get the relations
\[
    \frac{\partial \Psi_{1}}{\partial t_{k}} = k \phi_{k x}
\]
for \( k = 2, 3, \ldots \) besides \( \Psi_{1} = \phi_{1} \) that
follows from \eqref{eq:gen3.1}.

\subsection{$L^{-1}$ expansions for the twistor equations}
The twistor form for the factorization problem as given by eqs.
\eqref{eq:fac4} implies for the unknown function \( \phi(L, x) \)
the equations
\begin{equation}
    P(L,  x + \frac{\partial t}{\partial L} + \frac{\partial \phi}{\partial
    L} )\Big|_{-} = 0
    \label{eq:gen5}
\end{equation}
and
\begin{equation}
    X(L,  x + \frac{\partial t}{\partial L} + \frac{\partial \phi}{\partial
    L} )\Big|_{-} = 0.
    \label{eq:gen6}
\end{equation}
In order to obtain a solution \( \phi(L, x) \) for a fixed pair of
canonical variables \( (P, X) \) we should begin computing the
negative parts, as power series in the variable \( p \), of both
of the eqs. \eqref{eq:gen5} and \eqref{eq:gen6}. In this context
the following observation seems to be crucial in the whole
procedure of resolution developed in the sequel. Due to  the
connection \eqref{eq:gen2} between the variables \( p \) and \( L
\)  a negative power series in \( p \) can be written as a
negative power series in \( L \), since \( p^{-1} = L^{-1} (1 +
L^{-1} \phi_{x})^{-1} \), and reciprocally. The vanishing of the
negative parts of \( P \) and \( X \) as power series in \( p \)
is therefore equivalent to the vanishing the negative parts of \(
P \) and \( X \) as power series in \( L \). To take advantage of
the simplification gained in solving eqs. \eqref{eq:gen5} and
\eqref{eq:gen6} viewed as power series in \( L \), mainly because
they are naturally expressed in the variable \( L \) rather than
in \( p \), we should be able to compute the negative part of any
power of \( L \).


We shall presently proceed to compute the negative part of a power
series in the variable \( L \); i.e., its projection onto the
subalgebra \( \mathfrak{g}_{-} \) \textit{expressed in the
variable} \( L \). In that case we obtain for the projection of a
negative power of \( L \) the same negative power \( L^{-k}|_{-} =
L^{-k} \) for \( k = 1, 2, \ldots \) while for a positive power of
\( L \) we find the recurrent formula
\begin{equation}
    L^{k}|_{-} = [L^{k} - (L + \phi_{x})^{k}]_{-}
    \label{eq:neg1}
\end{equation}
for \( k = 0, 1, 2, \ldots \) This relation follows from the
identity \( L^{k}|_{-} = [L^{k} - p^{k}]_{-} \) after substitution
of \( p = L + \phi_{x} \) according to \eqref{eq:gen2}. To see why
this is a recurrent formula we develop the r.h.s. member according
to the binomial formula from which we get the desired relation
\[
    L^{k}|_{-} = -k L^{k-1} \phi_{x}|_{-} -\left( \begin{array}{c} k
    \\ 2 \end{array} \right) L^{k-2} \phi_{x} ^{2}|_{-} - \dots -
    \phi_{x} ^{k}.
\]

\subsection{Generating functions and symmetries}

As we said before, solutions of the factorization problem furnish
solutions of the dKP equation. The factorization problem is solved
once we know the function \( \psi_{-} \) or equivalently, as we have
just seen, the function \( \phi \) from which we obtain the dKP
solution \( u \) as
\begin{equation}
     u(x, t_{2}, t_{3}, \ldots) = -2 \phi_{1 x}(x, t_{2}, t_{3}, \ldots).
    \label{eq:gen4}
\end{equation}
This formula follows from the relation \( L^{2}|_{+} = p^{2} - 2
\phi_{1 x} = p^{2} + u  \) which is a direct consequence of
\eqref{eq:gen2} and the definition of \( L \).

The freedom allowed by the dKP equation for the definition of the
solution and  coordinates can be conveniently described in terms
of the action of the associated mechanical system. In
Hamilton--Jacobi theory, the zero-curvature equation in phase
space for the 1-form \( \omega_{+} = \sum_{n\geq 2} P_{n} \d t_{n}
\) defined by the Hamiltonians \( H_{n} = -P_{n} \), transforms in
the condition that the Poincar{\'e}--Cartan 1-form \( p \d x +
\omega_{+} \) be a closed form in configuration space. The action
\( S(x, t) \) is thereby locally defined according to the relation
\[
    \d S = p \d x + \omega_{+}
\]
and the Poisson structure disappears. In particular, the dKP equation
follows from
\begin{equation}
    \d S = p \d x + (p^{2} + u) \d y + (p^{3} +\frac{3}{2} u p + v) \d t.
   \label{eq:gen4.1}
\end{equation}
Incidentally, the function \( S(x, y, t) \) satisfies a modified dKP
equation
\[
    (S_{t} + \frac{1}{2} S_{x}^{3})_{x} = \frac{3}{2} S_{y} S_{xx} +
    \frac{3}{4} S_{yy},
\]
which transforms into the dKP equation through the Miura
map,
\[
    u = S_{y} - S_{x}^{2}
\]
as a consequence of \eqref{eq:gen4.1}. Going back to the definition of
coordinates, it is readily observed that \eqref{eq:gen4.1} is kept as
a closed form under the transformations:
\begin{align}
    x & =  \tilde{x} +\alpha (t), \quad y = \tilde{y}, \quad t = \tilde{t}
      \\
    \tilde{u}(\tilde{x},\tilde{y},\tilde{t}) & =  u(x, y, t) + \frac{2}{3}
    \dot{\alpha}
    \label{eq:gen4.2}
\end{align}
\begin{gather}
    x  =  \tilde{x} + \frac{2}{3}
    \dot{\beta} \tilde{y}
     \quad
    y  =   \tilde{y}+\beta, \quad t = \tilde{t}
      \\
    p  =  \tilde{p}+\frac{1}{3}
    \dot{\beta}
    \nonumber  \\
    \tilde{u}(\tilde{x},\tilde{y},\tilde{t})  =u(x, y, t) -\frac{2}{9}
    \dot{\beta}^{2}+\frac{4}{9}\ddot{\beta}\tilde{y}
    \label{eq:gen4.2.2}
\end{gather}
and
\begin{gather}
    x  =  \dot{\gamma}^{1/3} \tilde{x} + \frac{2}{9}
    \dot{\gamma}^{-2/3} \ddot{\gamma} \tilde{y}^{2}
     \quad
    y  =  \dot{\gamma}^{2/3} \tilde{y}, \quad t = \gamma(\tilde{t})
      \\
    p  =  \dot{\gamma}^{-1/3} (\tilde{p}-\frac{2}{9}
    \dot{\gamma}^{-1} \ddot{\gamma} \tilde{y} )
    \nonumber  \\
    \tilde{u}(\tilde{x},\tilde{y},\tilde{t})  =  \dot{\gamma}^{2/3}
    u(x, y, t) + \frac{2}{9} \dot{\gamma}^{-1} \ddot{\gamma} \tilde{x}
     + \frac{4}{27}(\dot{\gamma}^{-1} \gamma^{(3)} -
    \frac{4}{3} \dot{\gamma}^{-2} \ddot{\gamma}^{2}) \tilde{y}^{2}.
    \label{eq:gen4.3}
\end{gather}
This was  stated in \cite{14} where these symmetries were obtained
for the space-time metric used to derive the dKP equation.

\section{Reductions and the resolution of the the twistor equations}\label{sec:twi}

\subsection{Reductions}
To continue the analysis of the twistor system \eqref{eq:gen5},
\eqref{eq:gen6} we should make an explicit choice of the canonical
variables \( P,X \), which can be understood as reductions of the
general situation. In what follows we shall consider canonical
variables \( P, X \) of the form
\begin{equation}
    P(p, x) = \sum_{k\geq 0} a_{m-k}(\rho) p^{m-k}
    \label{eq:twi1}
\end{equation}
and
\begin{equation}
    X(p, x) = \sum_{k\geq 0} a_{n-k}(\rho) p^{n-k}
    \label{eq:twi2}
\end{equation}
where the power series depend on \( x \) through the new variable
\begin{equation}
    \rho := \frac{x}{h'(p)}, \hspace{.3in} h(p) := p^{r+1} e^{R_{-}(p)}
    \label{eq:twi2.1}
\end{equation}
defined in terms of the arbitrary negative power series in \( p \)
\[
    R_{-}(p) := \sum_{j\geq 1 }c_{j} p^{-j} .
\]
Some comments are in order about the definitions made above. The
structure of the transformation is fixed by a triple of positive
integers \( \{m, n, r\} \). The first two of them \( m, n \)
determine the positive degree in  \( p \) of the new variables \(
P = a_{m} p^{m} + \cdots \) and \( X = b_{n} p^{n} + \cdots \)
where the dots denote terms that contains of lower powers in \( p
\). The third integer \( r \) determines both the definition of
the variable \( \rho \) as well as the degree of the function \(
t(p) \) that we now take as a polynomial of degree \( r+1 \) in \(
p \),
\begin{equation}
    t(p) = t_{r+1} p^{r+1} + t_{r} p^{r} + \dots + t_{2} p^{2}.
    \label{eq:twi3}
\end{equation}
Regularity conditions on the coefficients \( a_{j}, b_{j} \) will
be assumed in each concrete case. At this point we shall postpone
the explicit construction of the canonical transformation \( (p,
x)\rightarrow (P, X) \) to the next section and concentrate in the
method for solving the twistor equations for the cases under
consideration.

The first step in that direction consists in the substitution in
\eqref{eq:twi1} and \eqref{eq:twi2} of \( (p, x) \) by \( (L, M) \) to
get new series in \( L \) that we now denote by
\begin{equation}
    F(L, x) := P(L, M) = \sum_{k\geq 0} a_{m-k}(\rho) L^{m-k}
    \label{eq:twi4}
\end{equation}
and
\begin{equation}
    G(L, x) := X(L, M) = \sum_{k\geq 0} b_{n-k}(\rho) L^{n-k}
    \label{eq:twi5}
\end{equation}
where the variable \( \rho \) is, of course, \( \rho = M/ h'(L) \)
for which, taking into account the expression \eqref{eq:gen3} for \(
M \), we find the formula
\[
    \rho = \frac{1}{h'(L)} \big(\frac{\partial t}{\partial L} + x + \phi_{L}\big)
\]
so that \( \rho \) is analytic in \( L \) at \( L = \infty \).
We assume generically that  the coefficients \( a_{j}(\rho) \) and \( b_{j}(\rho) \)
are analytic functions of \( L \) at \( L = \infty \). With that
hypothesis both \( F \) and \( G \) will continue having expressions
as power series in \( L \) with degrees \( m \) and \( n \) respectively.
The assumptions made in defining the canonical variables \( (P, X) \)
are enough to guarantee that the number of positive powers of \( L \)
appearing in \( F(L, x) \) and \( G(L, x) \) is in both cases finite.
This is an important requisite to carry out the present construction.

The structure of the system of equations \eqref{eq:gen5},
\eqref{eq:gen6} in the present situation follows from the analysis of
the series coefficients as we shall next show. Since it is assumed the
existence of power series expansions for the coefficients \(
a_{k}(\rho), b_{k}(\rho)  \), we have the series
\[
    a_{k}(\rho) = \sum_{j\geq 0} a_{k j}(x, t) L^{-j}, \hspace{.3in}
    b_{k}(\rho) = \sum_{j\geq 0} b_{k j}(x, t) L^{-j}
\]
where \( t = (t_{2}, t_{3},\ldots,t_{r+1}) \) represents the time
variables. It is easy to see that the coefficients in these series
are functions of the form
\[
    a_{k j}(x, t) = a_{k j}[x, t, \phi_{1},\dots,
    \phi_{j-r-1}],\hspace{.3in}
    b_{k j}(x, t) = b_{k j}[x, t, \phi_{1}, \dots,
    \phi_{j-r-1}]
\]
from which we deduce for \( F(L ,x) \) and \( G(L, x) \) in
\eqref{eq:twi4}, \eqref{eq:twi5} expansions of the form
\[
    F(L, x) = a_{m, 0} L^{m} + (a_{m, 1} + a_{m-1, 0}) L^{m-1}
    + (a_{m, 2} + a_{m-1, 1} + a_{m-2, 0}) L^{m-2} + \cdots
\]
and
\[
    G(L, x) = b_{n, 0} L^{n} + (b_{n, 1} + b_{n-1, 0}) L^{n-1}
    + (b_{n, 2} + b_{n-1, 1} + b_{n-2, 0}) L^{n-2} + \cdots
\]
that imply the expressions
\begin{equation}
    F(L, x) = \sum_{k\geq 0} f_{k} L^{m-k}, \hspace{.3in}
    G(L, x) = \sum_{k\geq 0} g_{k} L^{n-k}.
    \label{eq:twi6}
\end{equation}
We have thus obtained the formulas
\[
     f_{k}(x, t) = f_{k}[x, t, \phi_{1},\ldots,
    \phi_{k-r-1}],\hspace{.3in}
    g_{k}(x, t) = g_{k}[x, t, \phi_{1}, \ldots,
    \phi_{k-r-1}]
\]
displaying the number of functions \( \phi \) contained in each coefficient.
As in \eqref{eq:neg1} we continue with the notation \( F|_{-} \) and \( G|_{-} \)
for the projections on \( \mathit{g}_{-} \), the negative part, of the series \( F \) and \(
G \) of \eqref{eq:twi6}.
\begin{definition}\label{def:twi1}
    For the negative parts of the series \eqref{eq:twi6} we define
    the coefficients \( F_{k} \) and \(  G_{k} \) by the series
    \begin{equation}
     F|_{-} = \sum_{k\geq 1} F_{k} L^{-k}, \hspace{.3in}
     G|_{-} = \sum_{k\geq 1} G_{k} L^{-k}.
     \label{eq:twi7}
    \end{equation}
\end{definition}
 For instance, for \( F_{1} \) we find
\begin{equation}
    F_{1} = f_{0} L^{m}|_{-1} + f_{1} L^{m-1}|_{-1} + \dots + f_{m-1}
    L|_{-1} + f_{m+1}
    \label{eq:twi11}
\end{equation}
where \( L^{k}|_{-1} \) is the coefficient of \( L^{-1} \) in \(
L^{k}|_{-} \) as given by \eqref{eq:neg1}. We are now in a position to
formulate the main result of the present paper.

\begin{theorem}\label{thm:twi1}
    Assume the factorization problem \eqref{eq:fac1} has a solution
    described by the twistor equations corresponding to \eqref{eq:twi6}
   \begin{equation}
       F(L, x)|_{-} = 0, \hspace{.3in} G(L, x)|_{-} = 0
       \label{eq:twi8}
   \end{equation}
   for the canonical transformation defined by \eqref{eq:twi1},
   \eqref{eq:twi2}. Then, the solution \( u(x, t) = -2 \phi_{1 x} \),
   as given by \eqref{eq:gen4}, for the dKP
   hierarchy can be found solving for \( \phi_{1} \)
   the nonlinear system of \( m+n-2 \) ordinary differential equations
   for the  \( m+n-2 \) unknowns \( \phi_{1}, \phi_{2},\ldots, \phi_{m+n-2} \),
   \begin{equation}
       F_{1} =  F_{2} = \dots =  F_{n-1} = 0,
       \label{eq:twi9}
   \end{equation}
   \begin{equation}
       G_{1} =  G_{2} = \dots =  G_{m-1} = 0,
       \label{eq:twi10}
   \end{equation}
   determined by the coefficients in \eqref{eq:twi7}.
\end{theorem}

\begin{proof}
    Since the factorization problem is solvable there is a negative
    power series in \( L \), \( \phi(L, x) \), defining the canonical
    transformation \eqref{eq:gen1}, and being a solution of
    \eqref{eq:twi8}. As follows from the expression for \( L^{m}|_{-1}
    \) in \eqref{eq:neg1} this is a polynomial on the first \( m \)
    coefficients of \( \phi(L, x) \), \( \phi_{1x}, \phi_{2x},
    \ldots, \phi_{mx} \), hence the coefficient \( F_{1} \) of \( F
    \) in \( L^{-1} \) \eqref{eq:twi11} determines an equation of the form
    \[
        F_{1}(x, t, \phi_{1}, \ldots, \phi_{m-r}, \phi_{1x}, \ldots,
        \phi_{mx}) = 0
    \]
    where we have taken into account the structure of the coefficients
    \( f_{j} \) previously considered. By arguments of the same type
    one finds that \( F_{k} \) is of the form
    \[
        F_{k}(x, t, \phi_{1}, \ldots, \phi_{m+k-r-1}, \phi_{1x}, \ldots,
        \phi_{m+k-1, x}) = 0
    \]
    while for \( G \) we can write
    \[
        G_{k}(x, t, \phi_{1}, \ldots, \phi_{n+k-r-1}, \phi_{1x}, \ldots,
        \phi_{n+k-1, x}) = 0.
    \]
    Therefore, the first \( n-1 \) equations for \( F \), \eqref{eq:twi9}, and
    the first \( m-1 \) equations for \( G \),  \eqref{eq:twi10}, define
    a nonlinear system of first-order ordinary differential equations
    containing as many equations as unknowns and hence our assertion follows.
\end{proof}
In practice, for the examples that one can reasonably compute, things
become even simpler. The  sought function \( \phi_{1 x} \) can be
found \textit{algebraically} from  system \eqref{eq:twi9},  \eqref{eq:twi10}.
This is obviously always the case for \( r \geq m+n-2 \) but it needs
not to be so if \( r < m+n-2 \). An alternative interpretation of the
theorem is as a \textit{factorization criterion} for the problem
\eqref{eq:fac1}. For a given canonical transformation \( (p,
x)\rightarrow (P, X) \) one does not know in general whether there is a
solution for the system \eqref{eq:twi8}. But if a solution \( \phi_{1 x}
\) determined by \eqref{eq:twi9}, \eqref{eq:twi10} gives a solution \(
u = - 2  \phi_{1 x} \) of the dKP hierarchy, what one can check at
least for the dKP equation, then the whole \( \phi(L, x) \) can be
recovered. The question of whether or not the canonical transformation
is factorizable, in order to the system \eqref{eq:twi9},
\eqref{eq:twi10} admits a solution,  does not seem a serious
obstruction. For if we take transformations depending on arbitrary
functions and parameters generically enough, there will be special
values for the arbitrary data for which the solution ceases to exist.
In that case the solution \( u \) of the dKP equation becomes
singular for these special values of the free data.

\subsection{The canonical transformation and the reduction}\label{sec:can}

From what we have seen, one of the main ingredients in the whole
procedure allowing for the construction of solutions in the
twistor context is the canonical transformation \( (p,
x)\rightarrow (P, X) \) in Section \ref{sec:fac}. Thus we need a
description of the chosen canonical variables \eqref{eq:twi1},
\eqref{eq:twi2} which are determined by the differential equation
\( \d P \wedge \d X = \d p \wedge \d x \) that in terms of the
variables \( (p, \rho) \), where \( \rho \) is defined by
\eqref{eq:twi2.1}, becomes:
\[
    \d P \wedge \d X = h'(p) \d p \wedge \d \rho.
\]
Direct substitution of the series \eqref{eq:twi1}, \eqref{eq:twi2}
for \( P \) and \( X \) in this equation leads to a recurrent system
of ordinary differential equations for the coefficients \( b_{k}(\rho)
\) in terms of the arbitrarily given coefficients \( a_{k}(\rho) \) and
viceversa.

The most effective method of integration for these equations is
furnished by a generating function for the canonical transformation,
namely the function \( J(P, \rho) \) the differential of which is
given by
\[
    X \d P + h(p) \d\rho = \d J (P, \rho).
\]
The differential of this relation is the primitive equation for \( P \)
and \( X \) that will be identically fulfilled provided we have a
solution for the implicit equations
\begin{equation}
    h(p) = \frac{\partial J}{\partial \rho},
    \hspace{.3in} X = \frac{\partial J}{\partial P}
    \label{eq:can1}
\end{equation}
in terms of the arbitrary function \( J(P,\rho) \). The degree  of
arbitrariness for the generating function \( J(P,\rho) \) is
determined by the form of the new variables \( (P, X) \) fixed by
\eqref{eq:twi1}, \eqref{eq:twi2}. One should distinguish here
between two cases. For a given set of positive integers \( \{m, n,
r\} \) assume first we have \( m+n \geq r+2 \). In that case, it
is easy to see that we shall obtain the correct dependence of \( P
\) and \( X \) on the variables \( (p, \rho) \) if we define the
function \( J(P,\rho) \) according to the formula
\begin{equation}
    J(P,\rho) = \sum_{k = r+2}^{m+n} \gamma_{k} P^{\frac{k}{m}} +
    \sum_{k \geq 0} J_{r+1-k}(\rho) P^{\frac{r+1-k}{m}}.
    \label{eq:can2}
\end{equation}
Here the fractionary powers  are defined in terms of a fixed
$m$th-root of \( P \), the coefficients \( \gamma_{k} \) are
arbitrary constants with the restriction \( \gamma_{n+m} \neq 0 \)
for the coefficient of the leading term. Analogously, for the
arbitrary functions \( J_{k}(\rho) \) we impose the condition that
the derivative of the first term be different from zero, \(
J'_{r+1}(\rho) \neq 0 \). With these assumptions we shall prove
the existence of the announced canonical variables
\eqref{eq:twi1}, \eqref{eq:twi2}. Direct substitution of
\eqref{eq:can2} in equations \eqref{eq:can1} gives the equation
for \( P \)
\begin{equation}
    h(p) = \sum_{k \geq 0} J'_{r+1-k}(\rho) P^{\frac{r+1-k}{m}}
    \label{eq:can3}
\end{equation}
and also defines the variable \( X \) as
\begin{equation}
    X = \frac{1}{m} \sum_{k = r+2}^{m+n} k \gamma_{k} P^{\frac{k-m}{m}} +
   \frac{1}{m} \sum_{k \geq 0} (r+1-k) J_{r+1-k}(\rho) P^{\frac{r+1-m-k}{m}}.
    \label{eq:can4}
\end{equation}
In order to write the explicit form of the sought solution \( P \) to
\eqref{eq:can3} the variable \( P \) can be conveniently represented
as follows,
\[
    P(p, \rho) = a_{m}(\rho) p^{m} e^{A_{-}(p, \rho)},
\]
where \( A_{-}(p, \rho) \) is a negative power series in \( p \),
\[
    A_{-}(p, \rho) = \sum_{k \geq 1} A_{k}(\rho) p^{-k}.
\]
The coefficients \( a_{k}(\rho) \) for the series \eqref{eq:twi1}
defining \( P \) obtain from the series \(  A_{-}(p, \rho) \) which in
turn is computed through \eqref{eq:can3} according to the equation
\[
    e^{R_{-}} = J'_{r+1} a_{m}^{\frac{r+1}{m}} e^{\frac{r+1}{m}
    A_{-}} + J'_{r} a_{m}^{\frac{r}{m}} e^{\frac{r}{m} A_{-}} p^{-1}
    + \cdots
\]
The Taylor series in the variable \( \xi = p^{-1} \) at \( \xi = 0 \)
gives for the first coefficients the formulas
\[
    a_{m}(\rho) = \frac{1}{J'_{r+1}(\rho)^{m/r+1}}, \hspace{.3in}
    A_{1}(\rho) = \frac{m}{r+1}(c_{1}-\frac{ J'_{r}(\rho)}{
    J'_{r+1}(\rho)^{r/r+1} }),
\]
and so on. By substitution of the known \( P \) into the definition
\eqref{eq:can4} of \( X \) we get the final formula
\[
    X(p, \rho) = \frac{n+m}{m} \gamma_{n+m} (a_{m}(\rho))^{\frac{n}{m}} e^{\frac{n}{m}
    A_{-}(p, \rho)} p^{n} + \cdots
\]

As we said before, the structure of the canonical transformation
varies depending on the relative values of the integers in the set \(
\{m, n, r\} \). If instead \( m+n \geq r+2 \), the case we have just
considered, we would have \( m+n < r+2 \) then we should take for the
generating function \( J \) the simpler expression
\begin{equation}
    J(P,\rho) = \sum_{k \geq 0} J_{r+1-k}(\rho) P^{\frac{r+1-k}{m}}.
    \label{eq:can5}
\end{equation}
Then, it readily follows that the method given for computing the
variables \( (P, X) \) remains still valid for the new function \( J
\), but observe that \( n = r+1-m \).

In any case, although \( J \) is given by the series \eqref{eq:can2},
\eqref{eq:can5} it should be observed that only a finite number of terms
are enough to determine the solution of the dKP equation according to
eqs. \eqref{eq:twi9}, \eqref{eq:twi10} in theorem \ref{thm:twi1}. A
simple counting argument on the number of terms needed to find the
first \( m+n-1 \) coefficients in the series \eqref{eq:twi6} for \( F
\) and \( G \) leads to the following result.
\begin{theorem}\label{thm:can1}
    Let \( \{m, n, r\} \) a set of positive integers and \( (P, X) \)
    the corresponding canonical variables  \eqref{eq:twi1}, \eqref{eq:twi2}
    that define the twistor equations \eqref{eq:twi8} for the solutions
    of the dKP equation. Then, \( (P, X) \) can be found through
    \eqref{eq:can1} in terms of the generating function
    \[
        J(P,\rho) = \sum_{k = r+2}^{m+n} \gamma_{k} P^{\frac{k}{m}} +
    \sum_{k = 0}^{m+n-2} J_{r+1-k}(\rho) P^{\frac{r+1-k}{m}}
    \]
    if \( m+n\geq r+2 \), while for \( m+n < r+2 \) it is \( n = r+1-m
    \) and  we have
    \[
        J(P,\rho) = \sum_{k = 0}^{m+n-2} J_{r+1-k}(\rho) P^{\frac{r+1-k}{m}}.
    \]
\end{theorem}

\subsection{Explicit solutions}\label{sec:exp}

In this section we shall apply the previous results to the
construction of explicit solutions of the dKP equation \[
(u_{t}-\frac{3}{2} u u_{x})_{x} =\frac{3}{4} u_{yy} \] in the
simplest cases. To have a glimpse of the meaning of theorems
(\ref{thm:twi1}) and (\ref{thm:can1}) we shall analyze three
examples of increasing complexity.


\begin{enumerate}
\item When one of the canonical variables \( (P, X) \), say \( P
\), is of degree \( m = 1 \) then the solution to the dKP
hierarchy follows directly from \( F(L, x)|_{-} =0 \) in
\eqref{eq:twi9} independently of the concrete form of the
generating function \( J(P, \rho) \). In this case
        \[
        F(L, x) = a_{1}\big(\frac{\rho}{r+1}\big) L +
        a_{0}\big(\frac{\rho}{r+1}\big) + a_{-1}\big(\frac{\rho}{r+1}\big) L^{-1} +
        \cdots,
    \]
    and
    \[
        \rho = (r+1) t_{r+1} + r t_{r} L^{-1} + (r-1) t_{r-1}
        L^{-2} +\cdots
    \]
    The vanishing of the coefficient of \( L^{-1} \) leads to the
    equation,
    \begin{multline*}
        \frac{1}{2}[\frac{2(r-1)}{r+1}a'_{1}(t_{r+1}) t_{r-1} +
        (\frac{r}{r+1})^{2} a''_{1}(t_{r+1}) t_{r}^{2} ]  \\
        +\frac{r}{r+1} a'_{0}(t_{r+1} t_{r}) -a_{1}(t_{r+1}) \phi_{1 x}  =  0.
    \end{multline*}
    For the solution \( u = -2\phi_{1 x} \) we find the expression
    \[
        u = I_{2} t_{r-1} + \big(\frac{r^{2}}{2 (r^{2}-1)} I'_{2}-
        \frac{r^{2}}{4 (r-1)^{2}} I_{2}^{2}\big) t_{r}^{2} + I_{1} t_{r} +
        I_{0},
    \]
    where the arbitrary functions \( I_{j}(t_{r+1}) \) are an
    equivalent parametrization to that of the functions \( a_{j} (t_{r+1}) \)
    for the solution \( u \).

    Higher order coefficients \( \phi_{2 x}, \phi_{3 x}, \ldots \) can
be found from the equations corresponding to higher negative
powers of \( L \). In particular, for \( r = 2 \) it results the
solution of the dKP equation,
\[
    u(x, y,t) = I_{0}(t) + I_{1}(t) y + I_{2}(t) x + \big(\frac{2}{3}
    I'_{2}(t)-  I_{2}(t)^{2}\big) y^{2}.
\]
Observe that this solution is of a very simple nature. In fact, it
can be obtained from the zero solution $u=0$ by performing the two
symmetries of the dKP equation described in the previous section.

Consideration of higher values of \( m \) and \( n \) leads to
nice parametrizations for the solutions of nonlinear ordinary
differential equations. The parallelism with the theory of
algebraic curves for the KP equation, as a mater of fact, appears
manifestly in the present construction.

\item 
It is observed that for values of the set of integers \( \{m, n,
r\} \) in theorem (\ref{thm:can1}) which are \( \{2, 3, 2\} \) or
\( \{3, 2, 2\} \), \( \{3, 3, 2\} \), \( \{2, 4, 2\} \) or \( \{4,
2, 2\} \) the solutions of the dKP equation, as we shall see
below, are of the form \[ u = I + \sqrt{K} ,\] where \( I \) and
\( K \) represent polynomials in \( x, y \) with coefficients
depending on \( t \),
\[
    I(x, y, t) = I_{0}(t) + I_{1}(t) x +I_{2}(t) y + I_{3}(t) y^{2},
\]
and
\[
    K(x, y, t) = K_{0}(t) + K_{1}(t) x + K_{2}(t) y + K_{3}(t) y^{2}.
\]
If one introduces this form of the solution in the dKP equation
one gets the coefficients of \( I \) in terms of the coefficients
of \( K \), besides a Riccati equation for the coefficient \(
K_{3}(t) \), namely
\[
    K'_{3} = \frac{2}{K_{1}} K_{3}^{2} + \frac{16}{15}
    \frac{K'_{1}}{K_{1}} K_{3} + \frac{2}{15} K''_{1} -
    \frac{8}{45} \frac{(K'_{1})^{2}}{K_{1}}.
\]
A particular solution of this Riccati equation is
\[
K_3=\frac{2}{15}K'_1;
\]
hence, the general solution is easily found to be
\[
K_3=\frac{K_1^{8/5}}{C-2\int^t K_1(t)^{3/5} \d
t}+\frac{2}{15}K'_1.
\]

Finally, the formula for the solution of the dKP equation we were
looking for is:
\begin{align}
    u = & \frac{1}{ K_{1}^{2}} [-\frac{1}{2} K_{2}^{2} +
    \frac{2}{3} K_{1} K'_{0} - \frac{2}{9} K_{0}(3 K_{3} + 2 K'_{1})
  \\
     &+ \frac{2}{9}(K_{1} K'_{1} - 3 K_{1} K_{3}) x
   \\
     & +\frac{2}{9}(3 K_{1} K'_{2} - 2 K_{2} (6 K_{3} +
     K'_{1})) y
  \\
     &  +\frac{4}{3}(- K_{3}^{2} + \frac{1}{5} K_{3} K'_{1} - \frac{8}{90}
     (K'_{1})^{2} + \frac{2}{30} K_{1} K''_{1} ) y^{2}]
    \\
     &  +\sqrt{ K_{0} +  K_{1} x +  K_{2} y +
     K_{3} y^{2}}
    \label{eq:exp2}
\end{align}

This family of solutions lies in the orbit of a simpler solution
under the action of the previously mentioned symmetries of the dKP
equation. By these symmetries we can take $K_0=K_2=K_3=0$, so that
\[
\frac{2}{15} K''_{1} -
    \frac{8}{45} \frac{(K'_{1})^{2}}{K_{1}}=0.
\]
whose solution is
 \[
K_1(t)=\frac{1}{(at+b)^3},
  \]
  with $a$ and $b$ arbitrary constants.
  The corresponding solution of the dKP equation is
\[
   u := -\frac{2bx}{3(a+bt)}+\sqrt{\frac{x}{(a+bt)^3}}.
 \]
Observe that this is a a solution of the of stationary type
associated to the dKdV flow not depending on the $y$ variable.
%

We now proceed to give the $K$'s in some examples

\begin{enumerate}
\item $\{2,3,2\}$ We take in this case the complete form for $J$
and $h$:
\begin{align*}
J(P,\rho )&=P^2\gamma_4+P^{5/2} \gamma_5 +
      J_0(\rho) +P^{1/2} J_1(\rho)+ PJ_2(\rho) +
      P^{3/2}J_3(\rho),\\
h&=\exp(c_1p^{-1} + c_2p^{-2}+c_3p^{-3}+c_4p^{-4}) p^3
\end{align*}
and the corresponding $K$'s are
\begin{align*}
K_0=&-\frac{4 \dot J_3^{2/3}}{675\gamma_5^2}(-60c_1\gamma_5\dot
J_2\dot J_3^{1/3}\\&-9(3J_3^2-10\gamma _5(J_1-c_2t\dot
J_3^{1/3}))\dot
J_3^{2/3}+5c_1^2\gamma _5t(3\dot J_2+4t\ddot J_3)),\\
K_1=&-\frac{8\dot J_3^{5/3}}{15 \gamma _5},\\
K_2=&-\frac{16 \dot J_3^{2/3}}{45\gamma _5}(\dot J_2\dot
J_3^{1/3}-c_1 (\dot J_3+\frac{2}{3} t\ddot J_3)),\\
 K_3=&-\frac{16 \dot J_3^{2/3}\ddot J_3}{135\gamma _5}
\end{align*}
Observe that  $c_3,c_4,\gamma_4,J_0$ do not appear in the
solution.

\item $\{3,3,2\}$ While the general forms for $J$ and $h$ are:
\begin{align*}
J(P,\rho )&=P^{4/3}\gamma_4 + P^{5/3}\gamma_5 + P^2\gamma_6 \\&+
    J_{-1}(\rho)P^{-1/3}+J_0(\rho)+ J_1(\rho)P^{1/3}+J_2(\rho)P^{2/3}
    +J_3(\rho)P,\\
h&=\exp(c_1p^{-1} + c_2p^{-2}+c_3p^{-3}+c_4p^{-4}+c_5p^{-5}) p^3
\end{align*}
we again concentrate in the case when $c_4=c_5=\gamma_4=0$ and the
corresponding $K$'s are
\begin{align*}
K_0=&\frac{1}{45\gamma_5}(2 \dot J_3^{2/3}(-12 c_1t\dot J_2\dot
J_3^{1/3}+18 (J_1\dot J_3^{2/3}-c_2 t\dot J_3)+c_1^2t(3\dot
J_3+4t\ddot J_3)))
\\
K_1=&\frac{4\dot J_3^{5/3}}{5 \gamma _5},\\
K_2=&\frac{8 \dot J_3^{2/3}}{5\gamma _5}(\frac{1}{3}\dot J_2\dot
J_3^{1/3}-\frac{c_1}{3}
 (\dot J_3+\frac{2}{3} t\ddot J_3)),\\
 K_3=&\frac{8 \dot J_3^{2/3}\ddot J_3}{45\gamma _5}
\end{align*}
Observe that now $c_3,c_4,J_0$ do not appear in the solution.
\end{enumerate}

\item For different values of $m,n$ we get different type of
solutions, for example if we take $m=5,n=2$ and $J(P,\rho)= \gamma
P^{7/5}+\rho^2P^{3/5}$, $h=p^3$, we get the solution
\[
u=\frac{2}{2835\gamma^4t^2}\Big(\frac{A}{f}+B+C f\Big),
\]
where
\begin{align*}
A:=&(21)^{2/3}\gamma^7(-21870
(2)^{1/3}t^{22/3}+14\gamma(9tx-8y^2)^2),\\
B:=&7(9tx-8y^2)\gamma^4,\\
C:=&(21)^{1/3},\\
f:=&(24800580(2)^{1/3}t^{22/3}(2tx+y^2)\gamma^{11}+343\gamma^{12}(9tx-8y^2)^3\\&\,+
5(7)^{1/2}\gamma^{21/2}g^{1/2})^{1/3},\\
g:=&2510484768720\, t^{22}+2410616376 (2)^{2/3}\gamma
t^{44/3}(5751\,t^2x^2+5976\,txy^2+1394\, y^4)\\&\,+7715736
(2)^{1/3}\gamma^2t^{22/3}(9tx-8y^2)^3(27\,
tx+11\,y^2)+343\gamma^3(9tx-8y^2)^6.
\end{align*}
The same type is exhibited by the solution corresponding to
$m=5,n=2$ and $J(P,\rho)= \gamma P^{7/5}+\rho P^{3/5}$,
$h=p^3\exp(c/p)$
\[
u=\frac{1}{315\gamma^4}\Big(\frac{A}{f}+B+C f\Big),
\]
where now
\begin{align*}
A:=&(21)^{2/3}\gamma^7(-14 c^4\gamma+540\,t),\\
B:=&7c^2\gamma^4,\\
C:=&-(21)^{1/3},\\
f:=&(-343\gamma ^{11}c^6+5670\gamma^{11}(30x-20cy+11c^2t) +
5(7)^{1/2}\gamma^{21/2}g^{1/2})^{1/3},\\
g:=&18895680 t^3+183708\gamma
^7(900x^2-1200cxy+20c^2(33tx+20y^2)\\&\,-440c^3ty+113c^4t^2)-
15876c^6\gamma ^2(42x-28cy+13c^2t)+343 c^{12}\gamma ^3,
\end{align*}
\end{enumerate}

Cubic type solutions similar to those just presented here were
considered in \cite{20}.

\end{document}